# Elastic, electronic, optical and thermoelectric properties of $K_2Cu_2GeS_4$: a new chalcogenide material

M. A. Ali[a], M. A. Hossain[b], M. A. Rayhan[c], M. M. Hossain[a], M. M. Uddin[a], M. Roknuzzaman[d], K. Ostrikov[d], A. K. M. A. Islam[e,f], S. H. Naqib[f]

[a]Department of Physics, Chittagong University of Engineering and Technology (CUET), Chittagong-4349, Bangladesh.
[b]Department of Physics, Mawlana Bhashani Science and Technology University, Santosh, Tangail-1902, Bangladesh.
[c]Department of Arts & Science, Bangladesh Army University of Science and Technology, Saidpur-5310, Nilphamari, Bangladesh.
[d]School of Chemistry, Physics and Mechanical Engineering, Queensland University of Technology, QLD, 4000, Australia.
[e]Department of Electrical and Electronic Engineering, International Islamic University Chittagong, Kumira, Chittagong -4318, Bangladesh.
[f]Department of Physics, University of Rajshahi, Rajshahi-6205, Bangladesh.

## ABSTRACT

We report the first principles study of structural, elastic, electronic, optical and thermoelectric properties of newly synthesized $K_2Cu_2GeS_4$. The structural parameters are found to be in good agreement with experimental results. The single crystal elastic constants ($C_{ij}$) are calculated and $K_2Cu_2GeS_4$ is found to be mechanical stable. The analysis of polycrystalline elastic constants reveals that the compound is expected to be soft in nature. The values of Pugh and Poisson ratios suggested that the compound lies in the border line of ductile/brittle behavior. The chemical bonding is primarily ionic, the inter-atomic forces are central in nature and the compound is mechanically anisotropic. The computed electronic band profile shows semiconducting characteristics and the estimated band gap is strongly dependent on the functional used representing the exchange correlations. The nature of chemical bonding is explained using electronic charge density mapping. Important optical constants such as dielectric constants, refractive index, absorption coefficient, photoconductivity, reflectivity and loss function are

Corresponding Author: ashrafphy31@cuet.ac.bd

calculated and discussed in detail. Optical conductivity is found to be in good qualitative agreement with the results of band structure calculations. The Seebeck coefficients are positive for the entire temperature range used in this study, suggesting the presence of *p*-type charge carriers. We have obtained large Seebeck coefficent, 681 $\mu$V/K at 100 K and 286 $\mu$V/K at 300 K. At room temperature, the electrical conductivity and electronic thermal conductivity are $1.83\times10^{18}$ $(\Omega ms)^{-1}$ and $0.5\times10^{14}$ W/mK.s, respectively. The dimensionless figure of merit of $K_2Cu_2GeS_4$ is evaluated as ~1.0 at 300 K. This suggests that $K_2Cu_2GeS_4$ is a potential candidate for thermoelectric applications.

1.    **Introduction**

Over an extended period of time, the chalcogenides have gained considerable attention of the scientific community due to their fascinating structures as well as due to their prospect of excellent performance for photocatalysts, photoluminescence, photoresponse, nonlinear optics, topological insulators, and magneto-optic, magneto-ferroics and thermoelectric properties[1-9]. Quaternary sulfides with the formula of A/M/M'/S (A = alkali metal; M = group IB elements; M' = Ge, Sn), belong to the chalcogenides family, which exhibit semiconducting properties [10]. A number of Layered chalcogenides in the A-M-Sn-S system has been reported in literature [10-20]. For example, $Rb_2Cu_2SnS_4$ [18] and $K_2Ag_2SnSe_4$ [15] are synthesized by interesting layers $[Cu_2Sn_S4]^{2-}$ and $[Ag2SnSe4]^{2-}$, respectively, with defective anti-PbO-like structure. Only a few chalcogendes are reported in the class of A-M-Ge-S [10, 13, 17]. In addition, the tuning of physical properties of isostructural semiconductors is also expected due to element substitution [21]. For example, the band gap ($E_g$) of $K_4Cu_8Ge_3S_{12}$ (2.2 eV) [10] is larger than that of isostructural $K_4Cu_8Sn_3S_{12}$ (1.9 eV to 1.52 eV) [22]. $K_2Cu_2GeS_4$ is one of the chalcogenides belonging to the A-M-Ge-S system, very recently synthesized by Baohua Sun et al [21]. This material is very important because of their interesting crystal structure, tunable intermediate band gap and attractive optical properties. The crystal structure was found to be monoclinic (space group: P2/c). They [21] also studied electronic band structure and density of states of the new layered compound, $K_2Cu_2GeS_4$ and found that this compound is an indirect band gap (2.48 eV) semiconductor with interesting intermediate bands consisting of Ge 4$s$ and S 3$p$ states which reduced the band gap to 1.32 eV. However, any theoretical or experimental study on elastic, optical and thermoelectric properties of $K_2Cu_2GeS_4$ semiconductor are uninvestigated till date. The performance of a thermoelectric material is characterized by the so-called figure of merit, $ZT = \frac{S^2 \sigma T}{k}$, where $S$ is the Seebeck coefficient, $\sigma$ is the electrical conductivity, $k$ is the thermal conductivity, and $T$ is the absolute temperature. To realize an efficient energy conversion, the thermoelectric materials should have low thermal conductivity ($\kappa$), high electrical conductivity ($\sigma$), and large Seebeck coefficient ($S$). The low $\kappa$ is necessary to introduce a large temperature gradient between two ends of the material; whereas the high $S$ and $\sigma$ are needed to generate a high voltage per unit temperature gradient and to reduce the internal resistance of the material, respectively [23]. The presence of high electrical conductivity and low thermal conductivity in a single material is quite uncommon because both properties are positively correlated, in general. The semiconducting materials are potential candidates for thermoelectric devices which consist of $p$ and $n$-type materials [24]. Therefore, the searching of new layered chalcogenides semiconducting materials and study of their physical properties is of scientific interest from both research and application point of view. $K_2Cu_2GeS_4$ is composed of earth-abundant non-toxic materials which are preferable in thermoelectric device applications. Furthermore, the detail study of electronic and optical properties are also important for materials used in photocatalytic and photoelectric applications. These prospects motivate us to study the $K_2Cu_2GeS_4$ compound by means of first principles calculations. Therefore, an attempt has been made to study the structural, elastic, electronic, optical and thermoelectric properties of $K_2Cu_2GeS_4$, a new chalcogenide semiconducting material.

## 2. Methodology

The structural, elastic, electronic and optical properties of $K_2Cu_2GeS_4$ were calculated using the density functional theory (DFT) [25, 26] which is implemented in the Cambridge Sequential Total Energy Package (CASTEP) code [27]. The generalized gradient approximation (GGA) of the Perdew, Burke and Ernzerhof (PBE) [28] was adopted as the exchange and correlation terms. Besides, screened exchange local density approximation (sX-LDA) [29] was also used to calculate the electronic band structure. The electrostatic interaction between valence electron and ionic core was represented by the Vanderbilt-type ultra-soft pseudopotentials [30]. The cutoff energies of 400 eV were set for all calculations to ensure the precision. A *k*-point mesh of 4×5×3 according to the Monkhorst-Pack scheme [31] was used for integration over the first Brillouin zone. Broyden Fletcher Goldfarb Shanno (BFGS) geometry optimization [32] is used to optimize the atomic configuration. The thermoelectric properties, such as Seebeck coefficient, electrical conductivity, electronic thermal conductivity were calculated by solving Boltzmann semi-classical transport equations as implemented in BoltzTrap [33] interfaced with WIEN2k [34]. To obtain a good convergence, the basis set for self-consistent field (SCF) calculations a plane wave cut-off of kinetic energy $RK_{max} = 7.0$ was selected. A mesh of 416 *k*-points (13 × 16 × 8) in the irreducible representations were used for thermoelectric properties calculation. The generalized gradient approximation (GGA) within the Perdew-Burke-Ernzerhof (PBE) [28, 35] scheme was utilized for transport properties calculation. The muffin tin radii for K, Cu, Ge and S were fixed to 2.5, 2.28, 2.15 and 1.85 Bohr, respectively. The chemical potential was taken to a value which is equal to Fermi energy for transport properties calculation. The relaxation time $\tau$ was taken to be a constant. The electronic conductivity and the electronic part of thermal conductivity were calculated with respect to $\tau$, whereas the Seebeck coefficient was independent of.

**3. Results and Discussion**

3.1. Structural properties

The schematic of a unit cell of $K_2Cu_2GeS_4$ is shown in Fig. 1. The $K_2Cu_2GeS_4$ crystallizes in the monoclinic system with space group P2/c. The constituting atoms K, Cu, Ge and S are positioned by followings: $K_1$ at 2e; $K_2$ at 2a; Cu at 4g; Ge at 2f; $S_1$ at 4g and $S_2$ at 4g sites of Wyckoff coordinates, respectively, as shown in Table 1.

**Table 1** Optimized lattice constants (*a*, *b*, and *c*) and fractional coordinates of $K_2Cu_2GeS_4$.

|  | $a$ (Å) | $b$ (Å) | $c$ (Å) |  |  | Atomic position | | |
| --- | --- | --- | --- | --- | --- | --- | --- | --- |
|  |  |  |  | atoms | Sites | $x$ | $y$ | $z$ |
| $K_2Cu_2GeS_4$ | 7.118 | 5.388 | 11.212 | $K_1$ | 2e | 0 | -0.5267 | 0.25 |
|  |  |  |  | $K_2$ | 2a | 0 | 0 | 0.5 |
|  |  |  |  | Cu | 4g | -0.4997 | -0.7518 | 0.46756 |
|  |  |  |  | Ge | 2f | 0.5 | -0.2521 | 0.25 |
|  | 7.063[a] | 5.435[a] | 11.037[a] | $S_1$ | 4g | -0.2944 | -0.4702 | 0.4205 |
|  |  |  |  | $S_2$ | 4g | -0.2935 | -0.0340 | 0.1811 |

[a]Ref-[23]

The calculated (optimized) values of lattice parameters of $K_2Cu_2GeS_4$ are shown in Table 1 along with available experimental data [21]. Our calculated values are found in good agreement with the experimental values.

3.2. Elastic properties

3.2.1. *Single crystal elastic constants and mechanical stability:*

**Table 2** The calculated single crystal elastic constants $C_{ij}$ (in GPa) of $K_2Cu_2GeS_4$.

| $C_{11}$ | $C_{22}$ | $C_{33}$ | $C_{44}$ | $C_{55}$ | $C_{66}$ | $C_{12}$ | $C_{13}$ | $C_{15}$ | $C_{23}$ | $C_{25}$ | $C_{35}$ | $C_{46}$ |
|---|---|---|---|---|---|---|---|---|---|---|---|---|
| 53 | 52 | 57 | 32 | 10 | 17 | 18 | 13 | 6 | 30 | 6 | 5 | 0.25 |

In order to check the mechanical stability of the considered compound, we have calculated the single-crystal elastic constants using the finite strain technique. The response of a material under stress required to maintain a given deformation can be known from the elastic constants [36]. For a monoclinic crystal the criteria for mechanical stability are [37]: $C_{11} > 0$, $C_{22} > 0$, $C_{33} > 0$, $C_{44} > 0$, $C_{55} > 0$, $C_{66} > 0$, $[C_{11} + C_{22} + C_{33} + 2(C_{12} + C_{13} + C_{23})] > 0$, $(C_{33}C_{55} - C_{35}^2) > 0$, $(C_{44}C_{66} - C_{46}^2) > 0$, $(C_{22} + C_{33} - 2C_{23}) > 0$, $[C_{22}(C_{33}C_{55} - C_{35}^2) + 2(C_{23}C_{25}C_{35} - C_{23}^2C_{55} - C_{25}^2C_{33})] > 0$, $\{2[C_{15}C_{25}(C_{33}C_{12} - C_{13}C_{23}) + C_{15}C_{35}(C_{22}C_{13} - C_{12}C_{23}) + C_{25}C_{35}(C_{11}C_{23} - C_{12}C_{13})] - [C_{15}^2(C_{22}C_{33} - C_{23}^2) + C_{25}^2(C_{11}C_{33} - C_{13}^2) + C_{35}^2(C_{11}C_{22} - C_{12}^2)] + C_{55}(C_{11}C_{22}C_{33} - C_{11}C_{23}^2 - C_{22}C_{13}^2 - C_{33}C_{12}^2 + 2C_{12}C_{13}C_{23})\} > 0$. The calculated 13 single crystal elastic constants are given in Table 2. $K_2Cu_2GeS_4$ satisfies all the above criteria hence it is mechanically stable. In this system the largest component is $C_{33}$ and the second largest is $C_{11}$. The values $C_{11} \neq C_{22} \neq C_{33}$ [36] indicate the elastic anisotropy among the three principal axes.

3.2.2. Polycrystalline elastic constants

The polycrystalline elastic constants such as bulk modulus $B$, shear modulus $G$, Young's modulus $Y$ and Poisson's ratio $v$ can be calculated from these elastic stiffness moduli through the Voigt ($V$), Reuss ($R$) and Hill ($H$) approximations, [38-41] and the $V$ and $R$ approximations usually give the upper and lower bounds, respectively. In addition, the Young's modulus and Poisson's ratio can be obtained through the following equations: $Y = \frac{9BG}{3B+G}$, $v = \frac{3B-2G}{2(3B+G)}$ [42]. The calculated elastic moduli are presented in Table-3.

**Table-3** The bulk modulus ($B$), shear modulus ($G$), $G/B$ ratio, Young's modulus ($Y$), Poisson's ratio $v$, elastic shear anisotropy $A$ of $K_2Cu_2GeS_4$.

| $B$ (GPa) | $G$ (GPa) | $G/B$ | $Y$ (GPa) | $v$ | $A^U$ |
|---|---|---|---|---|---|
| 30 | 17 | 0.56 | 43 | 0.26 | 1.27 |

The bulk modulus and shear modulus can measure the response of the material to volume and shape change, respectively. Young's moduli can measure the resistance against uniaxial tensions. Although the parameters ($B$, $G$ and $Y$) do not measure directly the hardness of the materials but the values are normally large for hard substances. The values (Table 3) for the compound considered here, revealed that the material is very soft. Pugh [43] proposed a famous modulus ratio between $G$ and $B$, known as Pugh's ratio which separates the failure mode (ductility and brittleness) of a material. If the ratio is smaller (larger) than 0.57 for a material then it is said to be ductile (brittle). Table 3 shows the $G/B$ ratio which is very close (0.56) to 0.57 indicating that the compound studied here lies in the border line of ductile/brittle transition. In addition to Pugh ratio, Frantsevich's [44] also proposed a critical value of Poisson's ratio ($v \sim 0.26$) to separate the brittle and ductile nature of solids. The calculated value of $v$ ($v = 0.26$) is also demonstrating that the compound should lie in the border line of ductile/brittle in nature. The bonding nature can also be known from the value of *v* where the value of *v* is typically 0.10 for covalent bonding while it is 0.25 for ionic bonding in materials [45]. Therefore the chemical bonding is expected to be ionic within this compound. Moreover, the range of Poisson's ratio for central-force solids are 0.25−0.50 [46], thus, the value of *v* suggest that the inter-atomic forces are central. Since this is the first investigation of elastic properties of $K_2Cu_2GeS_4$, there is no experimental and theoretical data available and comparison is not possible at this time. It should be remembered that the errors in the calculated elastic constants are expected to lie between 5% to 15% [47].

In order to design a material with better mechanical durability it is very important to know elastic anisotropy of the material. It is often considered as one of the most decisive mechanical factors of a compound. Because, elastic anisotropy influences many physical properties such as development of plastic deformation in crystals, micro-scale cracking in ceramics, mechanical yield points, elastic instability, internal friction, etc. [48]. To study the anisotropy of a single crystal, we calculated the universal anisotropic index $A^U$, is defined as: $A^U = 5\frac{G_V}{G_R} + \frac{B_V}{B_R} - 6 \geq 0$, where $A^U = 0$ is for isotropic materials and the departure from zero defines the extent of anisotropy [49]. The value of $A^U$ is equal to 1.27 indicating the compound under consideration is anisotropic.

### 3.2.3. *Debye temperature*

The Debye temperature, $\Theta_D$ is an essential parameter of solids used to describe all the physical processes related to phonons; lattice vibration enthalpy, thermal conductivity, melting temperature, specific heat etc. The $\Theta_D$ can be calculated by a simple method using average sound velocity proposed by Anderson [50]. The calculation details could be obtained elsewhere [51-54]. The calculated values of density, sound velocities and $\Theta_D$ are presented in Table 4. As it is known, the harder is the solid, the higher is the Debye temperature. The calculated values of $B$, $G$ and $Y$ indicate the softness of $K_2Cu_2GeS_4$; hence a low value of $\Theta_D$ (373.24 K) is expected. Our calculated value of $\Theta_D$ can be compared with other chalcogenides $Cu_3TMS_4$ ($TM$ =V, Nb, Ta) [55] with $\Theta_D$ of about 356 K, 332 K, and 289 K, respectively.

**Table 4** Calculated density, longitudinal, transverse, mean sound velocities ($v_l$, $v_t$, and $v_m$) and Debye temperature $\Theta_D$ of $K_2Cu_2GeS_4$.

| $\rho$ (gm/cm$^3$) | $v_l$ (m/s) | $v_t$ (m/s) | $v_m$ (m/s) | $\Theta_D$ (K) |
|---|---|---|---|---|
| 1.70 | 5560.57 | 3159.19 | 3512.15 | 373.24 |

3.3. Electronic properties

The study of the electronic band structure is useful to explain many physical properties such as the optical spectra and transport properties of solids. As mentioned earlier that the electronic properties (band structure and density of states) has been studied [21]. Here a revisit on the band structure and density of states (DOS) has been made in order to assess the validity of our results with the available results. Moreover, we have also calculated the electronic charge density distribution. We have used two types of functionals to calculate the band structure, namely, GGA-PBE and sx-LDA. Generally, the GGA-PBE functional underestimates the interaction energy between valence electrons and the ions due to spreading of valence charge and hence underestimates the band gap. Underestimation of band gap of semiconductors is also previously reported [55-60]. Since the band gap of semiconductor is an important factor for practical application therefore calculation of band should more accurate. To do this calculation we have used sX-LDA functional and the calculated value of $E_g$ is found to be 2.27 eV which is in very good agreement with the experimental band gap (2.30 eV) [21]. The band structure has shown here only for GGA-PBE functional because the rest of the other properties presented here are calculated using this functional. The calculated electronic band structure is shown in Fig. 2 (a). The Fermi level ($E_F$) is set at 0 eV and coincides with the top of the valence band. It is clear from Fig. 2(a) that the valence band maximum (VBM) and the conduction band minimum (CBM) do not coincide each other indicating the indirect band gap nature with a band gap of about 1.37 eV GGA). Like Sun et al. [21], an intermediate band is also appeared at about 1.86 eV above the Fermi level and effective band gap is reduced further.

In order to explain the contribution from different atomic orbitals, we have also calculated the total and partial density of states (DOS). The wide valence band composed of three basic regions. (i) The lowest valence band (LVB) which is deeply bounded. The band is originated from the hybridization of S 3$s$, 3$p$ and Ge 4$s$ states. (ii) The mid valence band is considered within -6 eV to -2 eV energy range. The mid valence band originates from the hybridization of S 3$p$, Ge 4$p$ and Cu 3$d$ states with some but not dominant contribution from S 3$p$ states. (iii) The upper most loosely bound states are originated from predominantly Cu 3$d$ states and S 3$p$ states while the contribution from K 4$s$ & 3$p$ states is also noticeable but much smaller in magnitude. To explore the nature of chemical bonding in $K_2C_2GeS_4$ we have also calculated electronic charge density mapping (in the units of e/Å$^3$) along (101) crystallographic plane as shown in Fig. 3. The adjacent scale indicates the lower and upper values of the electronic density. Fig. 3 illustrates that there is a sharing of charge between Cu and S atoms forming covalent bond due to *pd* hybridization of Cu 3$d$ and S 3$p$ states below the Fermi level. The covalent bond is not very strong as was expected from the analysis of the elastic moduli. It is further noticed that a comparatively weaker covalent bond is also formed between S and Ge atoms due to *sp* hybridization below the Fermi level. Furthermore, metallic type bonds are supposed to exist in the Cu-Cu atoms. The bonding among K-S atoms is mainly ionic in nature.

3.4 Optical properties

The investigation of optical properties is very useful to understand the electronic structure of materials. To design the semiconductor optoelectronic devices, the refractive index and absorption coefficient should be well understood. The real and imaginary parts of dielectric function completely explain the optical properties of materials at all photon energies [61]. Therefore, calculations of optical properties are desirable to explore the potential application of materials.

The optical properties of $K_2Cu_2GeS_4$ have been studied via the frequency-dependent dielectric function $\varepsilon(\omega) = \varepsilon_1(\omega) + i\varepsilon_2(\omega)$ which is directly connected to the electronic structures of solids. The imaginary part, $\varepsilon_2$ of the dielectric function is calculated from the matrix elements between the occupied and unoccupied electronic states [62] and is given by

$$\varepsilon_2 = \frac{2e^2\pi}{\Omega\varepsilon_0} \sum_{k,v,c} |\psi_k^c|\boldsymbol{u}.\boldsymbol{r}|\psi_k^v|^2 \delta(E_k^c - E_k^v - E) \dots\dots\dots\dots\dots\dots\dots\dots (1)$$

where $u$ is used as the unit vector to describe the polarization of the incident electric field, $\omega$ is the frequency of the light, $e$ is the electronic charge, $\psi_k^c$ and $\psi_k^v$ are the wave functions of conduction and valence band, respectively at $k$, respectively. All other optical constants, such as refractive index, absorption spectrum, loss-function, reflectivity and conductivity (real part) can be calculated from the dielectric constants [62].

Fig. 4 shows the optical parameters for two polarization directions [100] and [001] up to energy range 0 to 20 eV. The two curves are almost identical in pattern; slight differences are in height of the peaks, indicating very small optical anisotropy. The electronic properties of crystalline materials are mainly characterized by the imaginary part, $\varepsilon_2(\omega)$ of dielectric function, $\varepsilon(\omega)$, which is related to the photon absorption phenomenon [63]. Fig. 4 (a) shows the real part of the dielectric function where the values of $\varepsilon_1(0)$ are found to be 4.35 and 5.65 for [100] and [001] direction, respectively indicating the dielectric nature of $K_2Cu_2GeS_4$. Fig. 4 (b) shows the imaginary part of the dielectric function in which the first critical point is observed at around 1 eV demonstrating the threshold for direct optical transition from valence band to the conduction band. The peaks in $\varepsilon_2(\omega)$ are associated with electron transitions. There are two peaks at around 4.6 eV and 6.8 eV for [100] direction and 3 eV and 6.5 eV for [001] direction. The peaks are also attributed to the electron transition from valence states to the conduction states.

The suitability of an optical material can be judged from the knowledge of the refractive index for its use in optical devices such as photonic crystals, waveguides etc [64]. The refractive index, *n* and extinction coefficient, *k* of complex refractive index of $K_2Cu_2GeS_4$ are displayed in Figs. 4 (c) and (d), respectively. The values of static refractive index are 2.08 and 2.32 for [100] and [001] direction, respectively. The refractive index curve is followed by real part of dielectric constant. On the other hand, the extinction coefficient is followed by imaginary part of dielectric function as it measures the absorption loss of electromagnetic radiation.

The absorption coefficient provides us with data about optimum solar energy conversion efficiency and it indicates how far light of a specific energy (wavelength) can penetrate into the material before being absorbed. Fig. 4(e) shows the absorption coefficient of of $K_2Cu_2GeS_4$ which starts rising at around 1.37 eV for [100] direction due to their semiconducting nature with a band gap of 1.369 eV. The curve for [001] direction is also started at around 1.32 eV. These are known as absorption edges. Interestingly, a strong absorption coefficient is observed in the region of 5.5 eV to 10.5 eV with a value greater than $10^5$ cm$^{-1}$. Fig. 4(f) shows that the photoconductivity does not start at zero photon energy due to the reason that the material has a distinct band gap which is also evident from the calculated band structure. When the incident photon energy is higher than that of the band gap, the photoconductivity starts. Moreover, the photoconductivity and hence electrical conductivity of the material increases as a result of absorbing photons [65]. The reflectivity spectra of K$_2$Cu$_2$GeS$_4$ as a function of photon energy are shown in Fig. 4 (g). The reflectivity spectra at two polarizations start from the zero frequency which is the static part of the reflectivity. The reflectivity for both directions is much lower in the visible and ultraviolet region which makes sure its potential applications in the area of transparent coatings in the visible and deep UV regions [66]. When a fast electron is moving through a material the energy loss of electron can be defined by a parameter known as loss-function. The peak in energy-loss function arises as $\varepsilon_1(\omega)$ goes through zero from below, $\varepsilon_2(\omega)$ is very small and abrupt reduction in reflectivity spectra occur. This particular value of energy is known as plasma frequency. The values of plasma frequencies are 11.3 eV and 11.7 eV for [100] and [001], respectively [Fig. 4 (h)].

3.5. Thermoelectric properties

The calculated Seebeck coefficient and the electrical conductivity of K$_2$Cu$_2$GeS$_4$ are plotted in Figs. 5. For a potential thermoelectric material, higher Seebeck coefficient and electrical conductivity are expected for higher power factor, but their coexistence is quite rare. The Seebeck coefficient is high (681 $\mu$V/K at 100K) and decreases rapidly up to 300 K and then the decrease rate become slower as shown in Fig. 5(a). The very small electrical conductivity at low temperature represent insulating behavior of K$_2$Cu$_2$GeS$_4$ as shown in Fig. 5(b) and these results are consistent with high Seebeck coefficient. The electrical conductivity increases with the increase in temperature and this once again reveal the semiconducting behavior of $K_2Cu_2GeS_4$. The thermoelectric figure of merit mainly depends on the Seebeck coefficient, $S = \frac{8}{3eh^2} \pi^2 k_B^2 m^* T (\frac{\pi}{3n})^{2/3}$ and electrical conductivity $\sigma = ne\mu$, here $k_B$, $h$, e, $T$, n, and m* are the Boltzmann constant, Planck constant, electronic charge, absolute temperature, carrier concentration, and carrier effective mass, respectively. The estimated Seebeck coefficient and electrical conductivity both satisfy the two conditions mentioned above. The Seebeck coefficient is positive for the entire temperature range suggesting the presence of p-type charge carriers in $K_2Cu_2GeS_4$. The room temperature thermoelectric properties of $K_2Cu_2GeS_4$ are presented in Table-5.

**Table-5:** Room temperature (300K) thermoelectric properties of $K_2Cu_2GeS_4$.

| $S$ ($\mu$V/K) | $\sigma/\tau$ ($10^{18}$/$\Omega$ms) | $S^2\sigma/\tau$ ($10^{10}$ W/mK$^2$s) | $\kappa_e/\tau$ ($10^{14}$ W/mK.s) | $ZT$ |
|---|---|---|---|---|
| 286 | 1.83 | 14.98 | 0.50 | 0.90 |

The temperature dependence of the power factor and electronic thermal conductivity are plotted in Fig. 6. Power factor is given by the combined effect of Seebeck coefficient and electrical conductivity. As the temperature increases power factor increases rapidly and after 600 K it increases slowly. The electronic thermal conductivity, is directly related with electrical conductivity ($k_e = L\sigma T$), increases with temperature in a very linear fashion after 300 K. The ultra-low electronic thermal conductivity is obtained in the range 100 to 300 K. At 300 K the $ZT$ value is 0.98 ~ 1.0 which is a promising for a potential thermoelectric material.

## 4. Concluding remarks

In summary, we have performed a first principles calculation for the structural, elastic, electronic, optical and thermoelectric properties of semiconducting $K_2Cu_2GeS_4$. The calculated single crystal elastic constants satisfy mechanical stability criteria. The values of *B*, *G*, and *Y* reveal that the compound is expected to be soft. The $K_2Cu_2GeS_4$ is also found to be elastically anisotropic. The calculated band structure implies $K_2Cu_2GeS_4$ an indirect band gap semiconductor. It is interesting that the hybridization of the intermediate bands due to Ge 4*s* and S 3*p* orbitals reduce the band gap. The band gap is calculated using both exchange correlation functional GGA-PBE and sX-LDA and the values are of 1.369 eV and 2.27 eV, respectively. From the analysis of electronic DOS and charge density mapping it is inferred that the chemical bonding in $K_2Cu_2GeS_4$ is expected to be a mixture of metallic, covalent and ionic contributions where ionic bonding is dominant. The dielectric nature of $K_2Cu_2GeS_4$ is expected from the value of static dielectric constant (4.35 and 5.65 for [100] and [001] direction, respectively). The calculated values of static refractive index are 2.08 and 2.32 for [100] and [001] direction, respectively. A good agreement is found among the optical and electronic properties such as electronic band gap (for A), absorption, and photoconductivity. The reflectivity spectra suggested that the semiconductor considered here might be used as transparent coating in the visible and deep UV region. Like elastic anisotropies, the optical properties are also found to be anisotropic with respect to different incident photon polarizations. The thermoelectric properties are calculated using BoltzTrap interfaced with Wien2k. We have obtained large Seebeck coefficient, 681 $\mu$V/K at 100 K and 286 $\mu$V/K at 300 K. At room temperature, the electrical conductivity and electronic thermal conductivity are $1.83 \times 10^{18}$ $(\Omega ms)^{-1}$ and $0.5 \times 10^{14}$ W/mK.s, respectively. The dimensionless figure of merit of *p*-type $K_2Cu_2GeS_4$ is evaluated as 0.98 at 300

K. Since our calculated Debye temperature for $K_2Cu_2GeS_4$ is relatively low, we expect low lattice thermal conductivity which should help to enhance the ZT value. Finally, we conclude that layered structured semiconducting $K_2Cu_2GeS_4$ is expected to be a potential candidate for thermoelectric device applications.

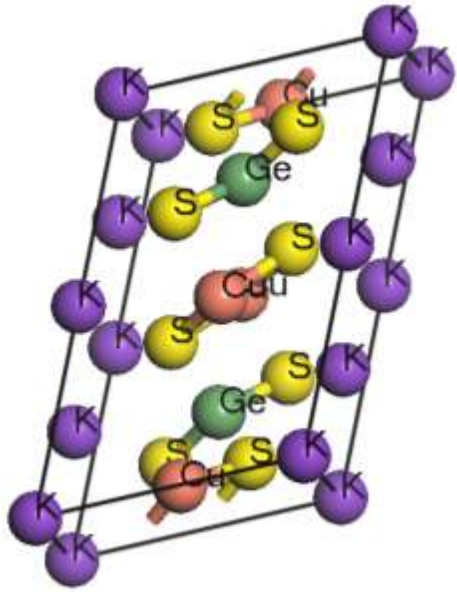

Fig. 1 Unit of cell of K$_2$Cu$_2$GeS$_4$

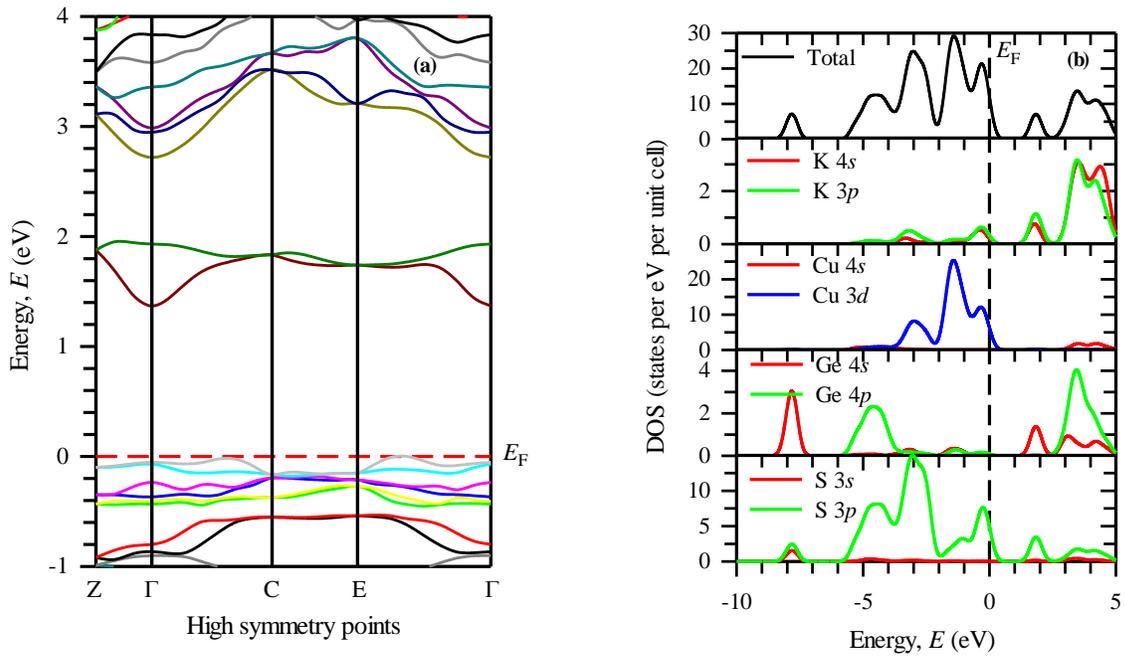

Fig. 2 (a) Electronic band structure and (b) density of states of K$_2$Cu$_2$GeS$_4$.

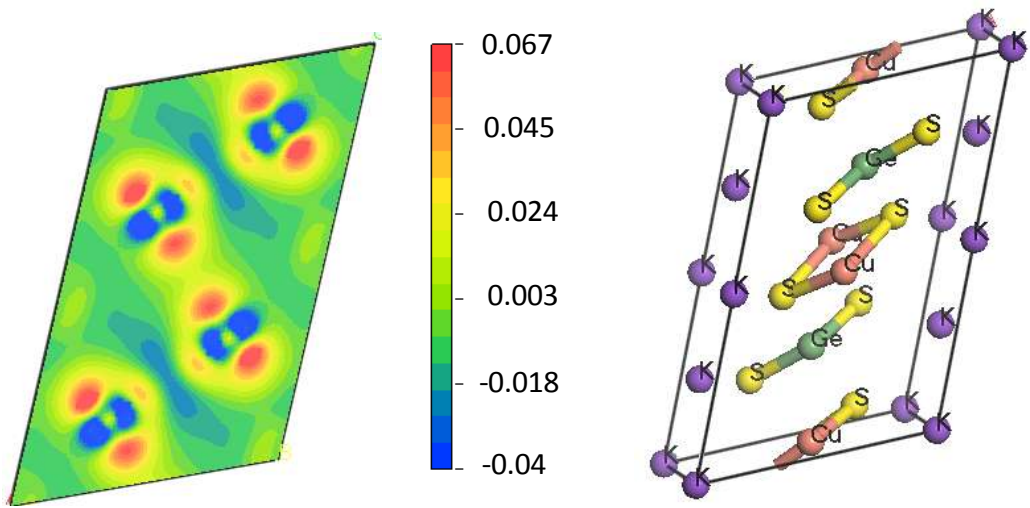

Fig. 3. Charge density distribution of $K_2Cu_2GeS_4$ along (101) planes

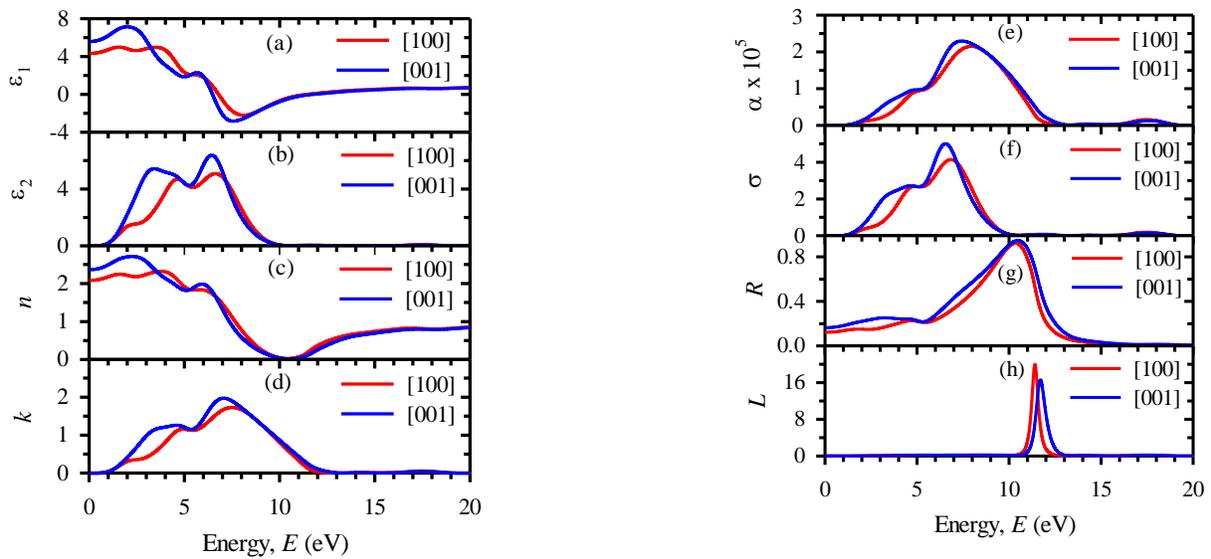

Fig. 4. Energy dependence of (a) real part of dielectric function, (b) imaginary part of dielectric function, (c) refractive index, (d) extinction coefficient, (e) absorption coefficient, (f) photo conductivity, (g) reflectivity, (h) loss function of $K_2Cu_2GeS_4$ for two polarization directions.

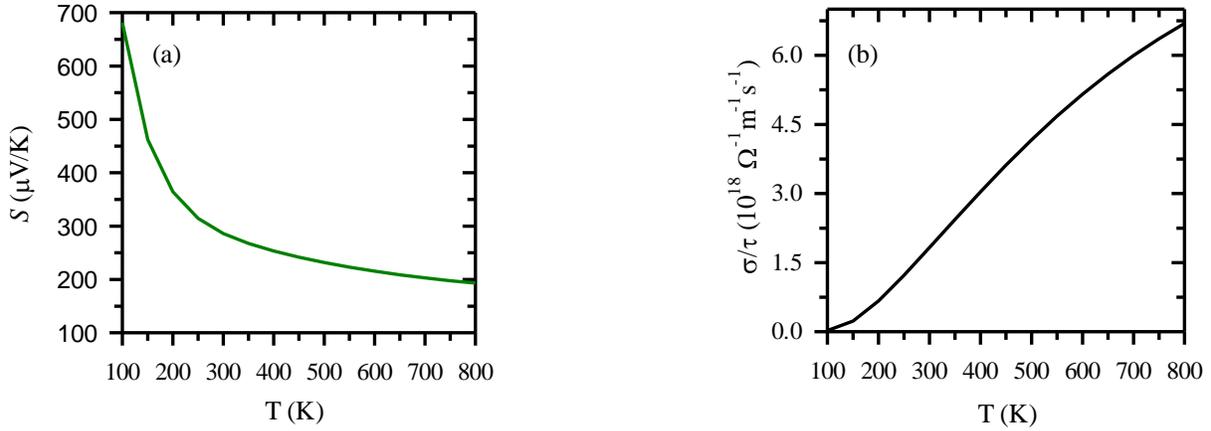

Fig. 5. Temperature dependence of (a) Seebeck coefficient and (b) electrical conductivity of semiconducting $K_2Cu_2GeS_4$.

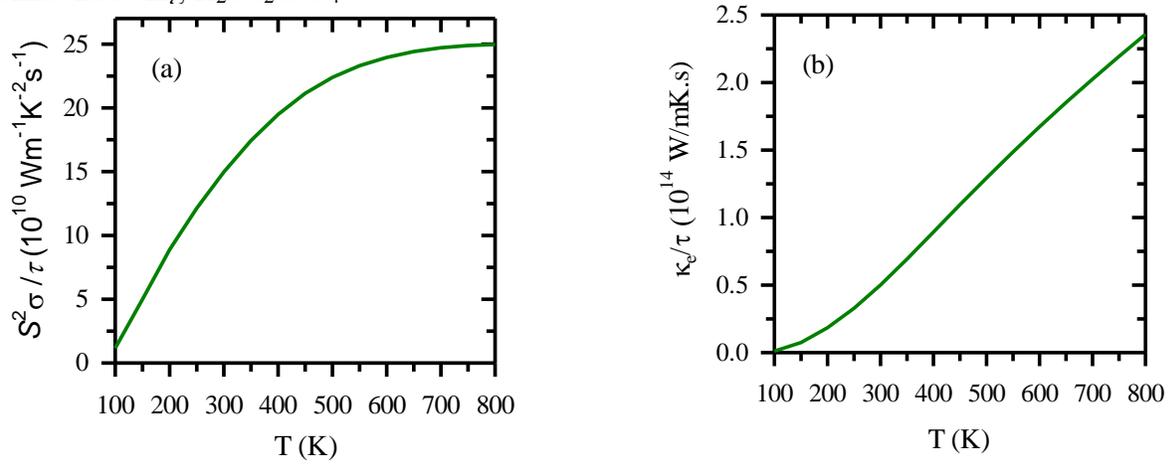

Fig. 6. Temperature dependence of (a) power factor and (b) electronic thermal conductivity of semiconducting $K_2Cu_2GeS_4$.

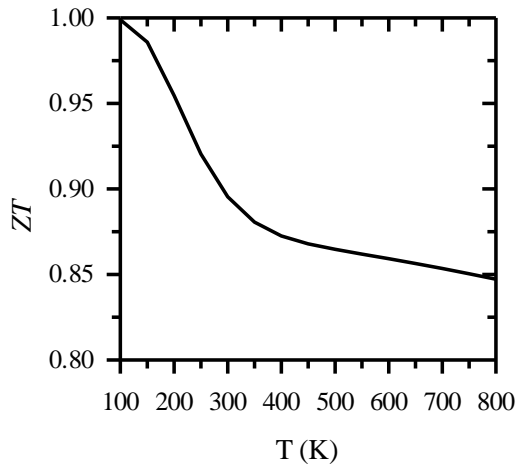

Fig. 7. Temperature dependence of figure of merit $ZT$ of $K_2Cu_2GeS_4$.